\documentclass[structabstract]{aa}  
\usepackage{graphicx}
\usepackage{txfonts}

\begin{document}

 \title{Density waves and star formation in grand design spirals}
 \author{Bernab\'e Cedr\'es
        \inst{1,2}
        \and
        Jordi Cepa
        \inst{1,2}
        \and
        \'Angel Bongiovanni
        \inst{1,2}
        \and
        H\'ector Casta\~neda
        \inst{3}
        \and
        Miguel S\'anchez-Portal
        \inst{4}
        \and
        Akihiko Tomita
        \inst{5}}
  \institute{Instituto de Astrof\'{\i}sica de Canarias (IAC), E-38200 La Laguna, Tenerife, Spain\\
  \email{bce@iac.es}
   \and
   Departamento de Astrof\'{\i}sica, Universidad de La Laguna (ULL), E-38205 La Laguna, Tenerife, Spain
   \and
   Departamento de F\'{\i}sica, Escuela Superior de F\'isica y Matem\'aticas, IPN, M\'exico D.F., M\'exico
   \and
   Herschel Science Centre, INSA/ESAC, Madrid, Spain
   \and
   Faculty of Education, Wakayama University, Wakayama 640-8510, Japan}
 
\date{Received; accepted}

 \abstract
{H~II regions in the arms of spiral galaxies are indicators of recent star-forming processes. They may have been caused by the passage of the density wave or simply created by other means near the arms. The study of these regions may give us clues to clarifying the controversy over the existence of a triggering scenario, as proposed in the density wave theory.}
{If passage of the density wave contributes significantly to the formation of H~II regions in the arms of grand design spirals, we can find a relationship in their position relative to the arms.} 
{Using H$\alpha$ direct imaging, we characterize the H~II regions from a sample of three grand design galaxies: NGC~5457, NGC~628 and NGC~6946. Broad band images in R and I were used to determine the position of the arms. The H~II regions found to be associated with arms were selected for the study. The age and the star formation rate of these H~II regions was obtained using measures on the H$\alpha$ line. The distance between the current position of the selected H~II regions and the position they would have if they had been created in the centre of the arm is calculated. A parameter, $T$, which measures whether a region was created in the arm or in the disc, is defined.}
{With the help of the $T$ parameter we determine that the majority of regions were formed some time after the passage of the density wave, with the regions located `behind the arm' (in the direction of the rotation of the galaxy) the zone they should have occupied had they been formed in the centre of the arm. Also, the relative efficiency of the star formation rate between arm and interarm was calculated for all the galaxies.}
{The presence of the large number of regions created after the passage of the arm may be explained by the effect of the density wave, which helps to create the star-forming regions after its passage. There is clear evidence of triggering for NGC~5457 and a co-rotation radius is proposed. A more modest triggering seems to exist for NGC~628 and non significant evidence of triggering are found for NGC~6946.}


\titlerunning{Density waves and star-forming regions}

\maketitle

\section{Introduction}
The density wave theory was established by the pioneering work of Lin \& Shu (1964, 1968) in order to explain the spiral structure present in many galaxies, even though there has been controversy with other mechanisms that have been proposed to explain the formation of the spiral arm (see, for example, Sellwood 2011 and references cited there;  Wada et al. 2011). Another controversy exists over whether the density wave is able to trigger star formation, as originally proposed by Roberts (1969), or whether it is just a means of reorganizing the material and the star-forming regions generated by other processes, such as the stochastic self-propagating star formation method proposed by Gerola \& Seiden (1978), which is much more prone to generating more fragmented, multiple arms.

Indeed, there exists clear evidence for the triggering of 
star formation in the arms of grand design galaxies. Cepa \& Beckman (1990) found that the star formation efficiency was higher in the arms that in the interarm regions of the galaxies NGC~3922 and NGC~628. Similar results were obtained by Knapen et al.\ (1996) for NGC~4321 and by Lord \& Young (1990) for M~51. Seigar \& James (2002) found, for a sample of 20 spiral galaxies, an increase in the H$\alpha$ flux near the K-band arms, and they interpreted this as star formation triggering by the density wave. Cedr\'es et al.\ (2005), using Monte Carlo simulations, found that there was a trend for the star-forming regions in a grand design galaxy to have an IMF biased towards a higher fraction of massive stars than the regions of a flocculent one. More recently, Grosb\o l \& Dottori (2009), using K-band images for the spiral galaxy NGC~2997, found that there was a trigger of star formation via the compression of the gas for the most massive regions, while smaller star-forming regions are formed more randomly over the rest of the disc.

On the other hand, the evidence for a non-triggering scenario is also compelling. Elmegreen \& Elmegreen (1984) found no evidence  for triggering  in a sample of 34 spiral galaxies when comparing the blueness of the arm with the interarm regions in the grand design galaxies. Elmegreen \& Elmegreen (1986), employing H$\alpha$ and UV bands, also found that the star formation rate (SFR) was independent of the presence of the density wave for a sample of galaxies.
Dobbs \& Pringle (2009), using a model of the interstellar medium and an estimate of the SFR based on the mass of gas in bound clumps, found no evidence of a dependence of the SFR on the strength of the spiral shock.
Recently, Foyle et al.\ (2010), using a multiband analysis, from H~I to the far-UV for the grand design galaxies NGC~5194, NGC~682 and NGC~6946, found that there is no `shock trigger', and that the spiral arms just reorganize the material from the disc, which will generate stars as a function of gas density, for example following a Kennicutt-Schmidt law (Kennicutt, 1998a).

With this study, we aim to determine whether the H~II regions associated by proximity with the arms of different grand design galaxies were formed by the influence of the density wave or were just associated with a higher density in the arm, and whether or not a triggering scenario exists. In section 2 we describe the selected galaxies and the data used. In section 3 we present the method developed and examine the dependence of the SFR on the position of the H~II region with respect to the arms. In section 4 we discuss the possible presence of triggering in the galaxies of the sample, and in section 5 a summary is given.

\section{Data}
We have selected for this study three grand design galaxies with low inclination (for a better identification of the spiral arms) and relatively close, in order to be able to gather data from a large number of H~II regions with good spatial resolution. To be able to trace the spiral arms more easily the galaxies had to be of grand design type. The galaxies selected were NGC~628 (M74), NGC~5457 (M101) and NGC~6946. All three galaxies have  arm class 9 according to Elmegreen \& Elmegreen (1987).

For NGC~5457 we took the data from Cedr\'es \& Cepa (2002). The catalogue presented in that paper contains data for 338 H~II regions in several emission lines, including the fluxes in H$\alpha$, information about the H$\alpha$ equivalent width, the size of the regions, and the extinction.
For the galaxies NGC~628 and NGC~6946 we employed data obtained through narrow band filter imaging from the Nordic Optical Telescope (NOT) with the ALFOSC instrument. These data, together with the reduction and calibration processes and a study of oxygen abundance, is presented in Cedr\'es et al.\ (2012).

For this study only the fluxes for H$\alpha$ line, the H$\alpha$ equivalent width and the size of the H~II regions are necessary.
In table \ref{param} there is a summary of the main parameters of the galaxies in the sample.

\begin{table*}
\caption{Parameters of the galaxy sample}
\label{param}
\begin{tabular}{c c c c c c c}
\hline\hline
Galaxy & $D$ (Mpc) & $\Omega_{P}$ (km/s/kpc) & P.A. (degrees) & $i$ (degrees) & 12+log(O/H)$_{\rm Central}$ & Rotation curve\\
\hline
NGC~628  & 7.3 (1) & 16$\pm$3 (3)  & 11.8 (3) & 6.5 (5) & 8.77 (6) & Helfer et al. (2003)\\
NGC~5457 & 7.2 (2) & 72$\pm$37 (3) & 38 (2)   & 18 (2) & 8.43 (7) & Sofue (1997)\\
NGC~6946 & 5.5 (2) & 39$\pm$8 (4)  & 64 (2)   & 30 (2) & 8.45 (6) & Sofue (1997)\\
\hline
\end{tabular}
\\
\tablefoot{(1) Helfer et al.\ (2003); (2) Sofue et al.\ (1999); (3) Lu et al.\ (1993); (4) Zimmer et al.\ (2004); (5) Kamphuis \& Briggs (1992); (6) Moustakas et al.\ (2010); (7) Cedr\'es et al.\ (2004)}

\end{table*}

\section{H~II region parameters}

\subsection{Arm characterization and selection}
The arms of the galaxies of the sample were defined  using images in the I-band obtained from the literature (Knapen et al.\ 2004). The width of the arms for each galaxy was calculated as follows: the I-band image was smoothed using a median filter in order to wipe out the high spatial frequency variations in flux of the disc and arms of the galaxy. This filter was a square one with a size of 10$\times$10 pixels (approximately twice the mean seeing value).
Then, several radial profiles, from the center of the galaxy (but avoiding the nucleus) that intersected the arms, were generated. The contribution from the disc was removed by fitting an exponential to the brightness profile of the galaxy (Freeman 1970), and then this exponential was subtracted from the data.  
Gaussians were then fitted to the residuals in the places where the arms were intersected. We can then assume that the FWHM of the fitted Gaussians is the width of the arms. For NGC~628 and NGC~5457 the median value of the width is 0.5 arcmin. For NGC~6946 is 0.54 arcmin.

The arm positions were traced by determining the maximum emission point in boxes large enough to contain the full width of the arm in the zone but small enough to be unaffected by the other arms, stars and/or bright HII regions. The typical size of the boxes was about 0.7$^{\prime}\times$0.7$^{\prime}$, which is equivalent to 1.5$\times$1.5 kpc$^2$ for NGC~5457 and NGC~628, and 1.1$\times$1.1 kpc$^2$ for NGC~6946.

\begin{figure}
\centering
\resizebox{\hsize}{!}{\includegraphics{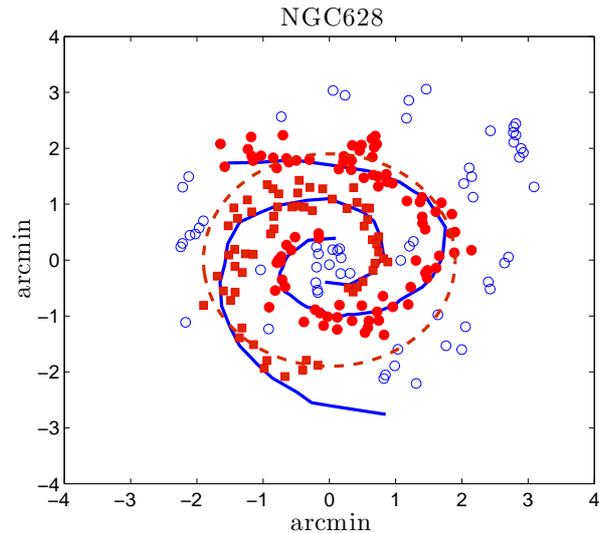}}
  \caption{Position of the HII regions detected in the galaxy NGC~628. Filled dots are the regions considered to belong to the NSN arm, while the filled squares mark the regions considered to belong to the SNS arm. The dashed line represent the position of the co-rotation. North is at the top and east is at left.}

  \label{arm628}
\end{figure}

\begin{figure}
\centering
\resizebox{\hsize}{!}{\includegraphics{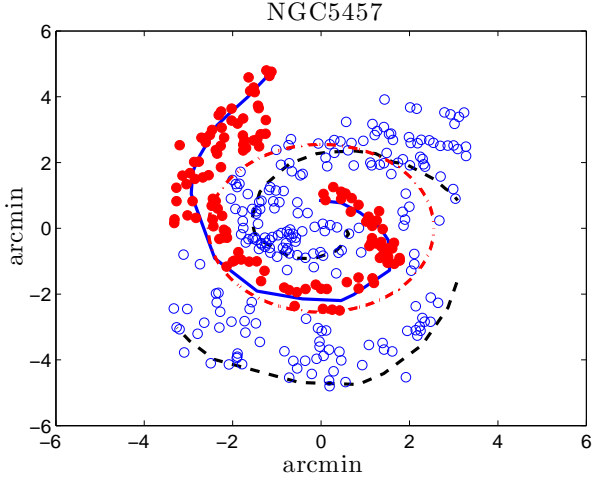}}
  \caption{Position of HII regions and the arms in the galaxy NGC~5457 from Cedr\'es \& Cepa (2002). The filled dots are the regions selected for this study and that possibly belong to the arm marked with the continuous line. The dashed lines represent the other arms detected for this galaxy. The dot-dashed line represent the possible location of the co-rotation radius (see section 4.2). North is at the top and east is at left.}
  \label{armm101}
\end{figure}

\begin{figure}
\centering
\resizebox{\hsize}{!}{\includegraphics{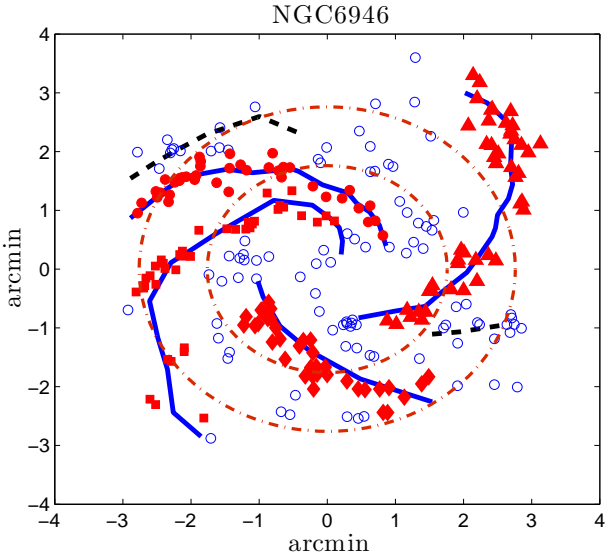}}
  \caption{Position of the HII regions detected in the galaxy NGC~6946. Filled dots are the regions considered for the main north arm. Filled squares, filled diamonds and filled triangles are the regions considered for the NS arm, south arm and SN arm respectively. Two secondary satellite arms, not included in the study, are represented by the dashed lines. The inner dot-dashed line is the $\Omega-\kappa/3$ resonance, and the outer dot-dashed line is the $\Omega-\kappa/4$ resonance. North is at the top and east is at left.}
  \label{arm6946}
\end{figure}

The regions were associated with each arm using a proximity criterion: they were assumed to belong to the arm they were closer to when they were inside (totally or partially) the width of the arm, and there were no uncertainties concerning their association to a satellite arm or any other kind of structure not directly related to the main arm. Following this criterion, nuclear/circumnuclear regions were also not considered.
For NGC~628 only the two main arms will be included in the study (Fig. \ref{arm628}).
In the case of NGC~5457, its complex structure, with several discontinuous satellite arms, makes it feasible to study only the NSN arm. In Fig. \ref{armm101}  the inner parts of this galaxy with the HII regions detected by Cedr\'es \& Cepa (2002) and the positions of the central part of the arms are indicated.
For NGC~6946, four arms will be considered, as indicated in Figure \ref{arm6946}.

In table \ref{regions} a summary of the arms employed for each galaxy and the number of regions associated with each arm is presented. 


\begin{table}
\caption{Number of regions in each of the arms selected for the galaxies in the sample}
\label{regions}
\centering
\begin{tabular}{c c}
\hline\hline
NGC~628 &  \\
\hline
Arm NSN & 86 regions\\
Arm SNS & 60 regions\\
Total (all arms \& interarm) & 209 regions\\
\\
\hline\hline
NGC~5457 & \\
\hline
Arm NSN & 122 regions\\
Total (all arms \& interarm) & 338 regions\\
\\
\hline\hline
NGC~6946 & \\
\hline
North Arm& 30 regions\\
Arm NS & 43 regions\\
South Arm& 34 regions\\
Arm SN & 35 regions\\
Total (all arms \& interarm) & 226 regions\\
\end{tabular}
\end{table}

\subsection{Age determination}
\begin{figure}
 \centering
 \resizebox{\hsize}{!}{\includegraphics{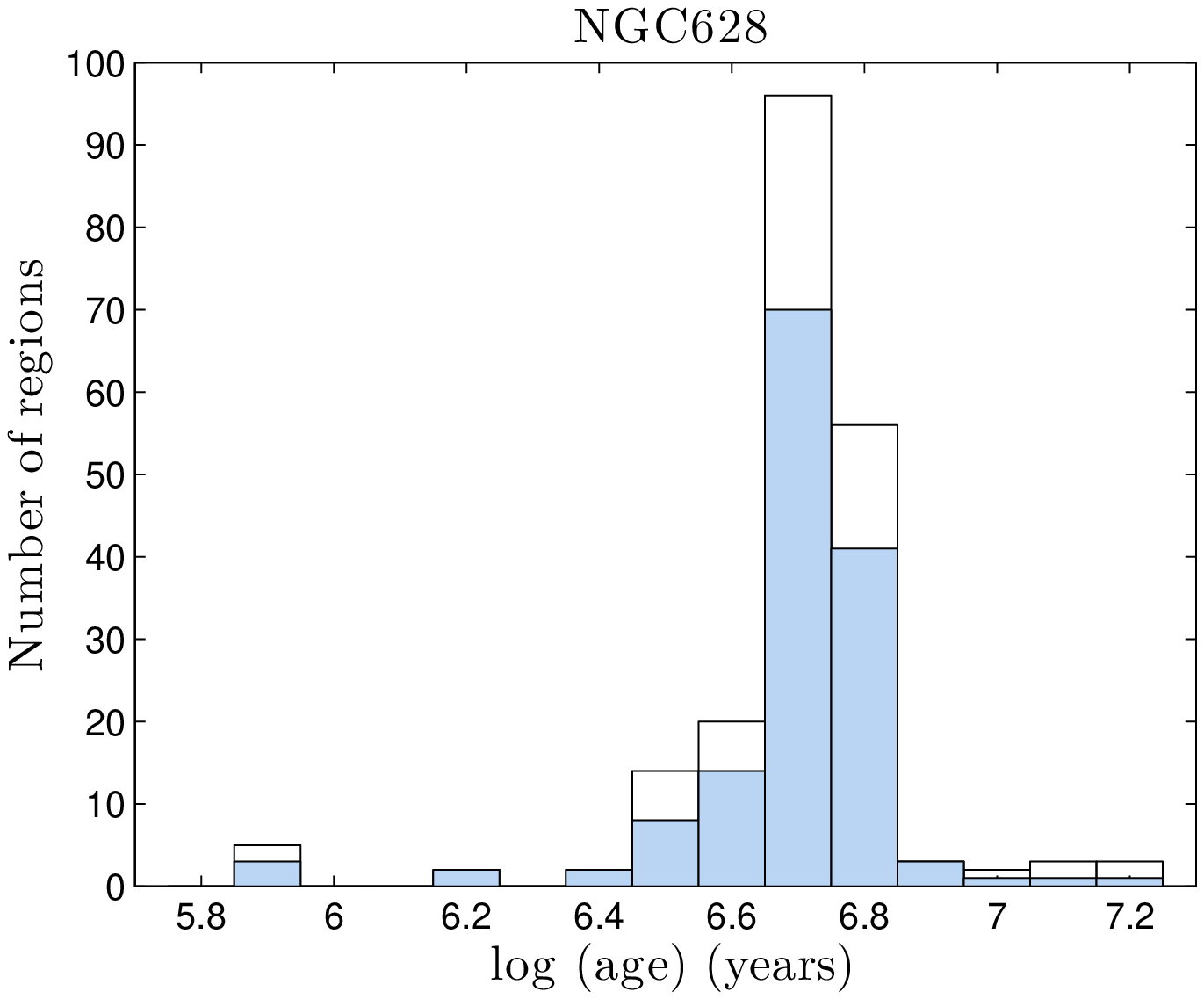}}
 \caption{Age histogram for NGC628. The open bars represent the whole sample of H~II regions for this galaxy. The shaded bars are the regions selected for the study in the arms as indicated in Figure \ref{arm628}.}
 \label{age62}
\end{figure}

\begin{figure}
 \centering
 \resizebox{\hsize}{!}{\includegraphics{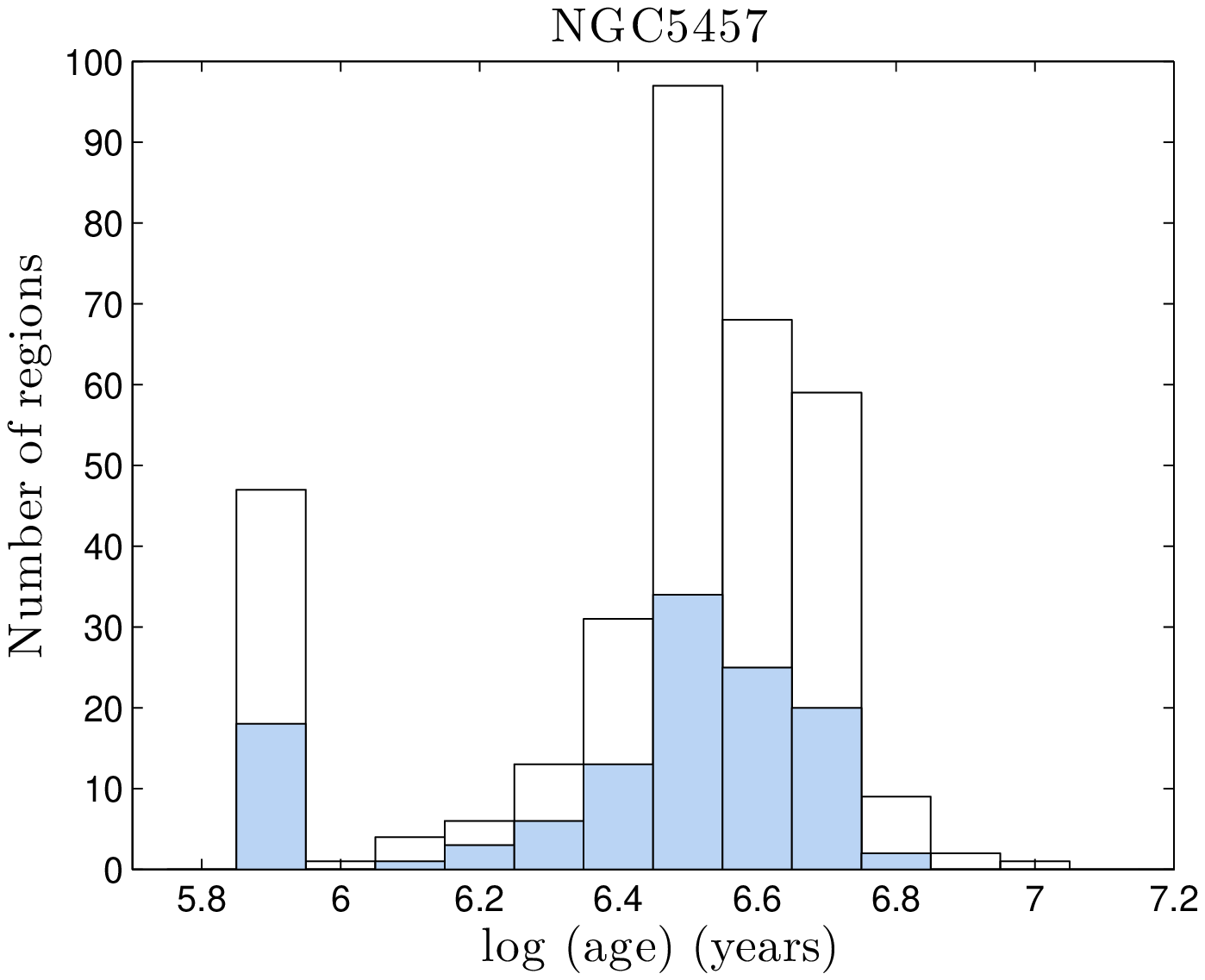}}
 \caption{Age histogram for NGC5457. The open bars represent the whole sample of H~II regions for this galaxy. The shaded bars are the regions selected for the study in the arm as indicated in Figure \ref{armm101}.}
 \label{age54}
\end{figure}

\begin{figure}
 \centering
 \resizebox{\hsize}{!}{\includegraphics{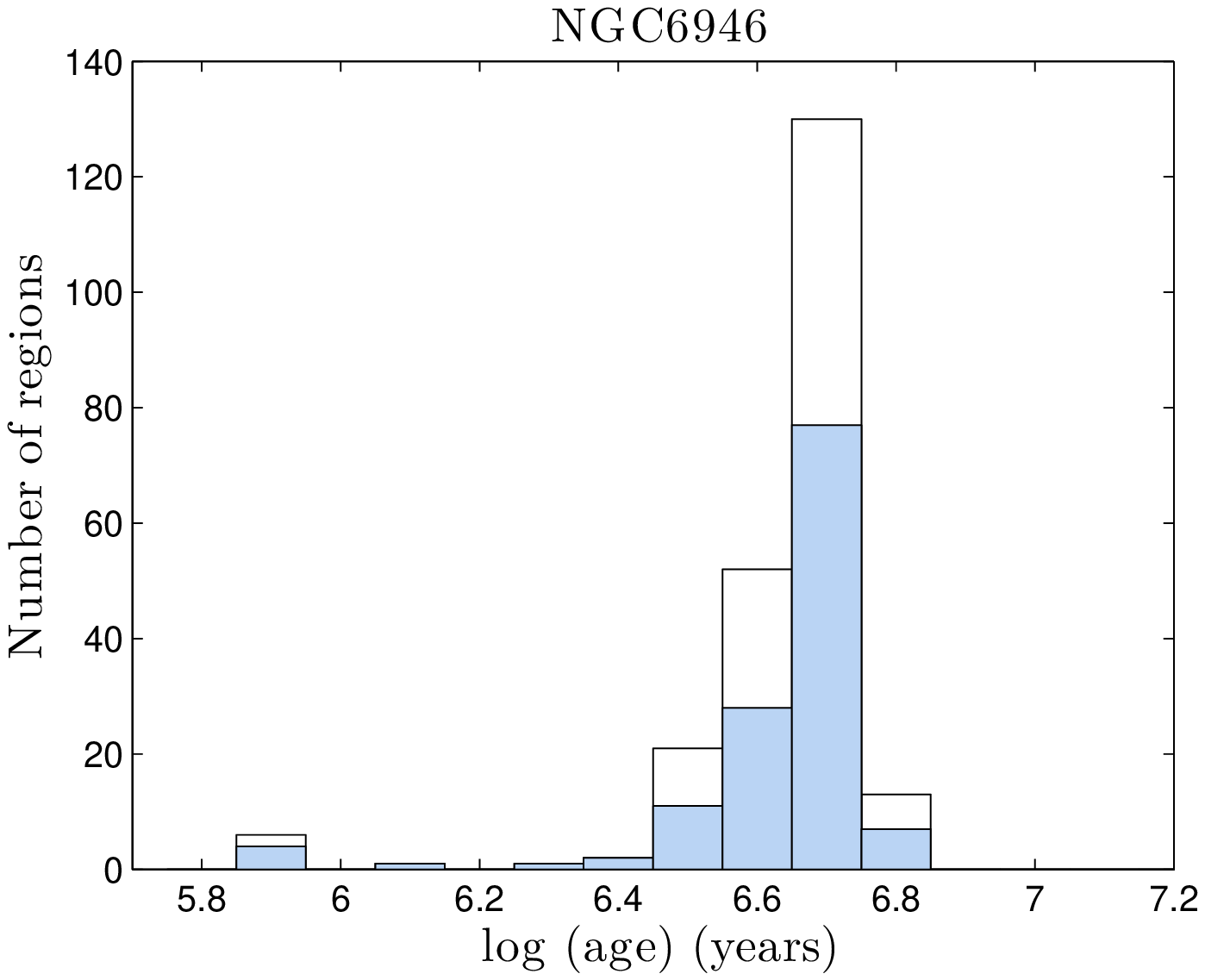}}
 \caption{Age histogram for NGC6946. The open bars represent the whole sample of H~II regions for this galaxy. The shaded bars are the regions selected for the study in the arms as indicated in Figure \ref{arm6946}.}
 \label{age69}
\end{figure}

\begin{figure}
 \centering
 \resizebox{\hsize}{!}{\includegraphics{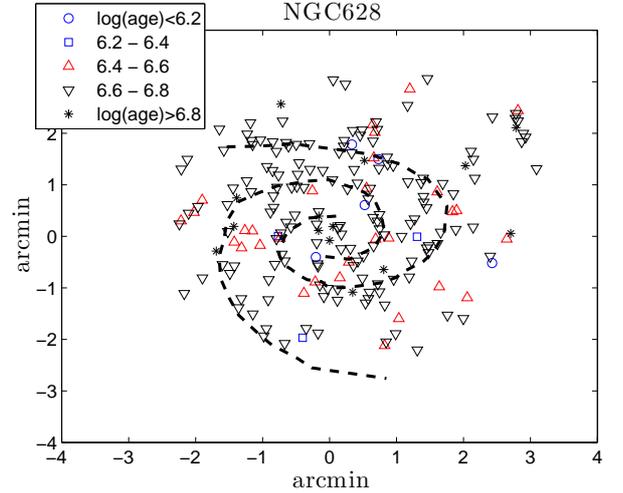}}
 \caption{Age map for the H~II regions of NGC628. The meaning of the different symbols is as indicated.} 
 \label{agem62}
\end{figure}

\begin{figure}
 \centering
 \resizebox{\hsize}{!}{\includegraphics{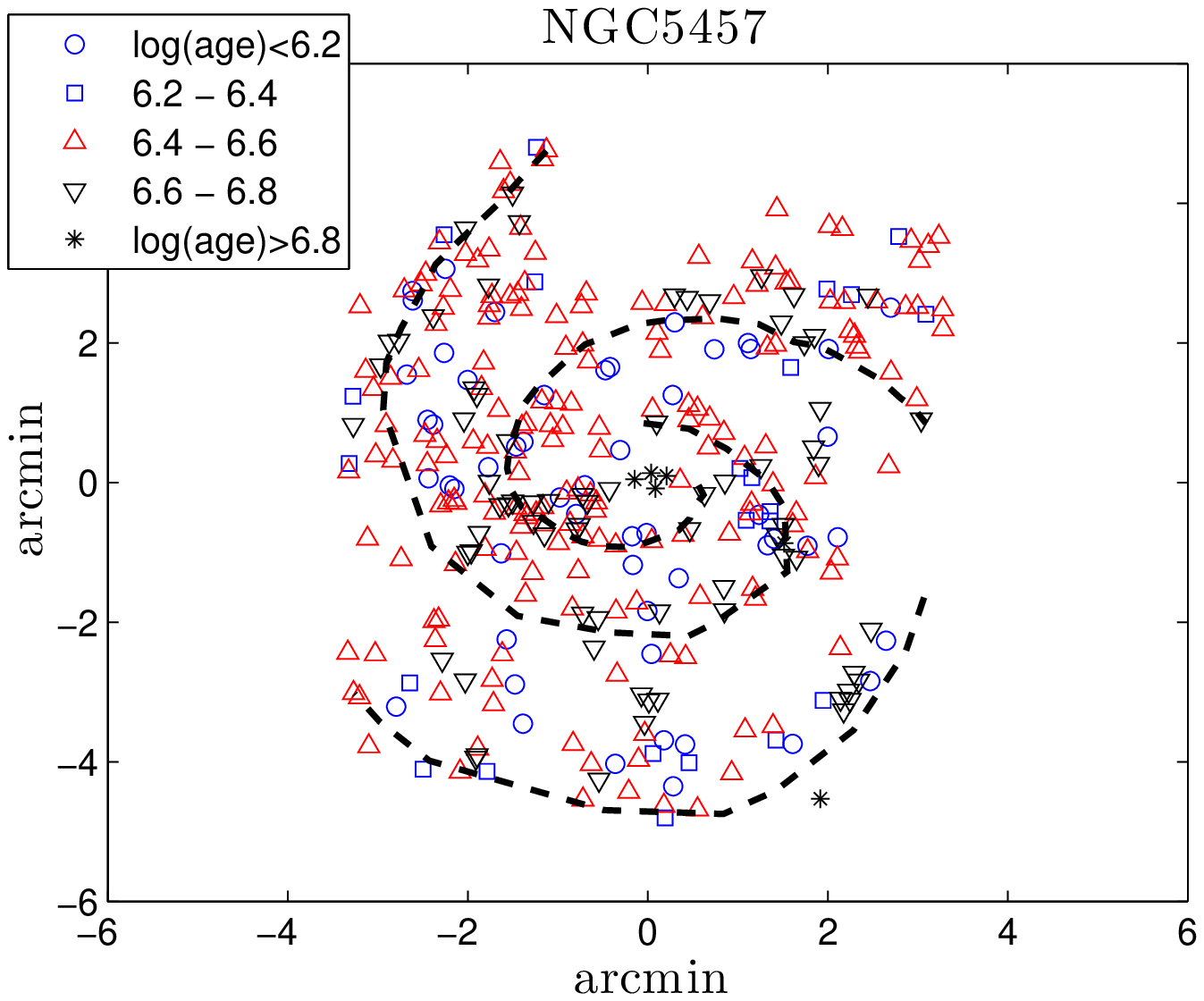}}
 \caption{Age map for the H~II regions of NGC5457. Meaning of the symbols as in Figure \ref{agem62}.}
 \label{agem54}
\end{figure}

\begin{figure}
 \centering
 \resizebox{\hsize}{!}{\includegraphics{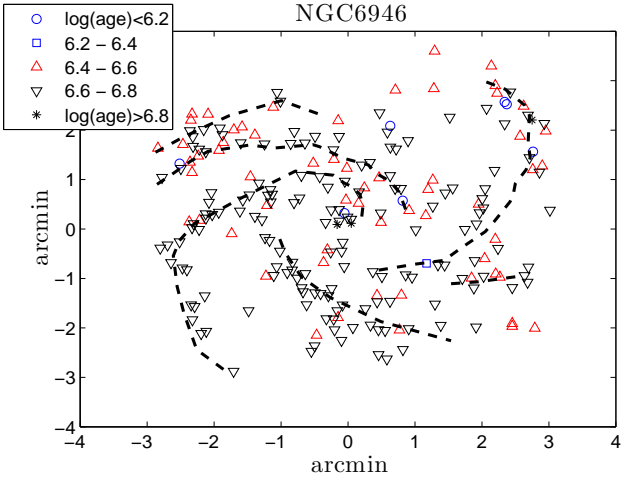}}
 \caption{Age map for the H~II regions of NGC6946. Meaning of the symbols as in Figure \ref{agem62}.}
 \label{agem69}
\end{figure}

The age of the HII regions of the galaxies was obtained using the H$\alpha$ equivalent width through Leitherer et al.\ (1999) models. 
These models present three different initial mass functions (IMFs) available: a Salpeter function (1995), with a slope of -2.35 and an upper mass of 100\,M$_{\odot}$, a truncated Salpeter, with an upper mass of 30\,M$_{\odot}$, and an IMF with a slope of -3.30 and an upper mass of 100\,M$_{\odot}$. The variation of the derived ages due to the use of different IMFs is less than 3\% in most cases, which is smaller than the uncertainties introduced by the errors in the equivalent width that dominates the determination of the age of the regions. Therefore, we simply assumed a Salpeter (1995) IMF as the most probable one for our H~II regions. We also assumed an instantaneous burst for the star formation rate.
Five metallicities are available from Leitherer et al.\ (1999) models: $Z=$ 0.040, 0.020 ($Z_{\odot}$), 0.008, 0.004 and 0.001, where $Z=Z_{\odot}\times10^{(\log(O/H)-12+3.1)}$. From table \ref{param}, the abundances are, without considering the gradients, between $Z\simeq$ 0.008 and $Z=Z_{\odot}$. However, as with the determination of the IMF, the uncertainties derived from the use of different abundances are much smaller than the uncertainties that introduce the errors in the equivalent widths, so the results in the determination of the age of the regions are almost independent of the metallicity range in which the H~II regions are located. For this reason we selected only the metallicity $Z=Z_{\odot}=0.020$.
For the regions with equivalent widths larger that the highest value tabulated for the model, a minimum age of $\log (age)=5.9$ (with the age expressed in years) was assigned.
In Figs \ref{age62}, \ref{age54} and \ref{age69} the age histograms for all the detected regions of NGC628, NGC5457 and NGC6946, respectively, are represented (the regions selected for this study are marked by the shaded bars).
In Figs \ref{agem62}, \ref{agem54} and \ref{agem69} the age maps for NGC~628, NGC~5457 and NGC~6946 are represented in the same order.

\begin{table}
\caption{Median ages of all the H~II regions of the galaxies used in this study.}
\label{ages}
\centering
\begin{tabular}{c c c}
\hline\hline
Galaxy & Median age (Myr) & Median absolute deviation (Myr)\\
\hline
NGC~628 & 5.37 & 1.32 \\
NGC~5457 & 3.38 & 1.49 \\
NGC~6946 & 5.07 & 0.86 \\
\hline\hline
\end{tabular} 
\end{table}


Table \ref{ages} shows the median value of the distribution of ages obtained from all the H~II regions of the galaxies in the sample and represented by the histograms in Figures \ref{age62} to \ref{age69}.
From this table and Figs \ref{age62} to \ref{agem69}, even if it seems that the HII regions of NGC~5457 are, in general terms, $\simeq$2\,Myr younger than the regions from NGC~628 and $\simeq$1.6\,Myr younger than the regions from NGC~6946, the large value of the dispersion of the data does not allow us to conclude that such an age difference is significant. However, this result for NGC~5457 is in good agreement with those determined by Egusa et al.\ (2009), and our general results for NGC~628 are in the same range of those obtained in the study by S\'anchez-Gil et al.\ (2011), where a different method for determining the age of the star-forming regions was used (H$\alpha$ to far-UV flux ratio).

\subsection{Definition of the $T$ parameter}
\begin{figure}
  \centering
  \resizebox{\hsize}{!}{\includegraphics{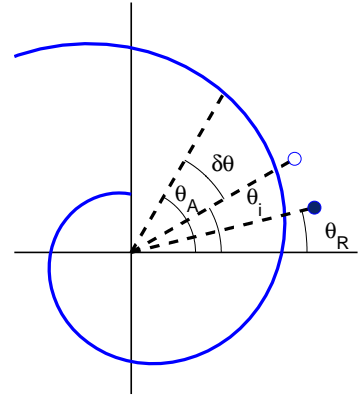}}
  \caption{Explanation of the deduction of the offset between the hypothetical position of an H~II region formed in the centre of the arm and its real position for a clockwise rotating galaxy. The trailing spiral arm is represented by the thick line. The hypothetical H~II region is represented by the open circle. The real position of the H~II region is indicated by the filled circle. The angles $\theta_{i}$, $\theta_{R}$, $\theta_A$ and the offset $\delta\theta$ (between the arm and the theoretical position of the H~II region) are indicated by  an arc and the appropriated label. In this diagram it is assumed that the H~II region is inside the co-rotation radius of the galaxy.}
  \label{spiral}
\end{figure}

Assuming that an H~II region of age $t_i$ was formed in the centre of the arm, its hypothetical position can be expressed in polar coordinates on the plane of the galaxy disc as $(\theta_{i},R_{i})$, where $\theta_i$ is the angle of the region (measured counterclockwise) and $R_{i}$ is its galactocentric radius. These coordinates were obtained from the observed one (i.e. measured over the tangent plane of the sky) by de-projection with respect to the line of sight using the data in Table \ref{param}. In the same way, we can describe the observed position of the arm where this region was supposedly formed by the coordinates $(\theta_A$, $R_A)$, where $\theta_A$ is the angle of the centre of the arm and $R_A$ its radius.
Taking into account that the radial velocities of the H~II regions are an order of magnitude lower than the circular velocities of the star-forming regions in the density wave model (Roberts 1969; Kuno \& Nakai 1997), we may assume that the H~II region has suffered no significant net movements, so that $R_i=R_A$. In Fig. \ref{spiral} there is schematic drawing of a clockwise rotating arm where the position of the angles are indicated.

Following Egusa et al.\ (2009), the angular offset between this hypothetical H~II region and the centre of the arm can be expressed as
\begin{equation}
\delta\theta=\theta_{A}-\theta_{i}=(\Omega-\Omega_{p} ) \times t_{i},
\end{equation}
where $\Omega$ is the rotation velocity of the galaxy at this radius and $\Omega_{p}$ is the pattern speed associated to the density wave (see Table \ref{param} for the rotation curves and the $\Omega_p$ employed in this study).
We can consider now the real position of an H~II region, which is described by the coordinates $(\theta_{R},R_{R})$. Again, if we assume that the region has suffered no radial displacements, then $R_R=R_A=R_i$.
The offset between the position of a real H~II region and the hypothetical one can be expressed as
\begin{equation}
\Delta\theta=\theta_{i}-\theta_{R}
\end{equation}
This can be rearranged as (see Fig. \ref{spiral})
\begin{equation}
\Delta\theta=\theta_{A}-(\delta\theta+\theta_{R})
\label{deltateta}
\end{equation}
Then, $\Delta\theta$ gives us an indication of how far an H~II region is located from its expected angular position if it had been formed in the centre of the arm. If the angles are measured in the direction of the disc rotation, the value of $\Delta\theta$ is positive when the region is `ahead' of its hypothetical position and negative when it is `behind'. This result does not depend on whether the H~II region is located inside or outside the co-rotation radius.

However two regions with the same $\Delta\theta$ would have different linear distances to the arm depending on the radial position of the region, so $\Delta\theta$ is not a good parameter to quantify the distance to the density wave. Moreover, the linear distance obtained with $\Delta\theta$ will also be dependent on the radius through the differential rotation of the galaxy disc, so in order to compare regions at different radii the following parameter is used:

\begin{equation}
\left(\frac{T}{Myr}\right)=17.0648\left(\frac{\Delta\theta}{\circ}\right)\left(\frac{km/s/kpc}{\Omega}\right),
\label{parat}
\end{equation}
where $T$ is the time required for the region to move the $\Delta\theta$ angle. This parameter is independent of the position of the region in the galaxy.
A positive value of $T$ (or $\Delta\theta$), means that the distance from the centre of the arm to the position of the H~II region ($\theta_A - \theta_R$) could not have been traveled by this region in is lifetime; therefore, the region was not formed in the centre of the arm, but before the passage of the arm over the molecular cloud in which the H~II region will form (the previously defined `ahead' position). Vice versa, a negative value of $T$ (or $\Delta\theta$) means that the region was formed after the passage of the arm (the `behind' position).

\begin{figure}
  \centering
  \resizebox{\hsize}{!}{\includegraphics{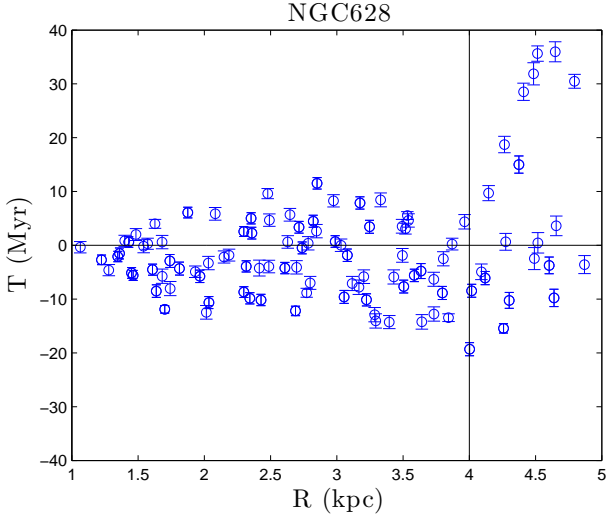}}
  \caption{Parameter $T$ (Eq.\ \ref{parat}) as a function of galactocentric distance for the sample of selected HII regions (arm regions) of NGC628.}
  \label{rt62}
\end{figure}

\begin{figure}
  \centering
  \resizebox{\hsize}{!}{\includegraphics{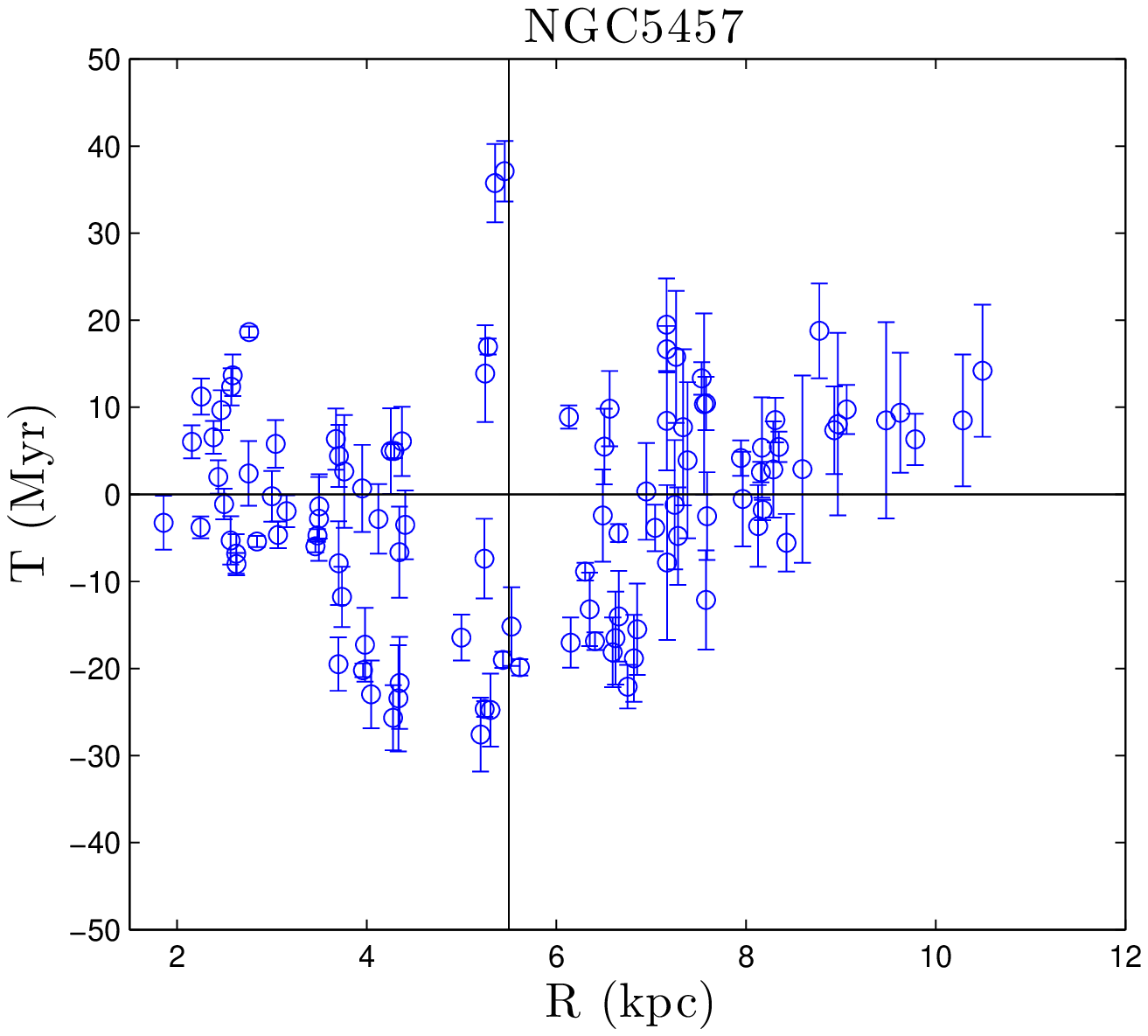}}
  \caption{Parameter $T$ (Eq. \ref{parat}) as a function of galactocentric distance for the sample of selected HII regions (arm regions) of NGC5457.}
  \label{rt54}
\end{figure}

\begin{figure}
  \centering
  \resizebox{\hsize}{!}{\includegraphics{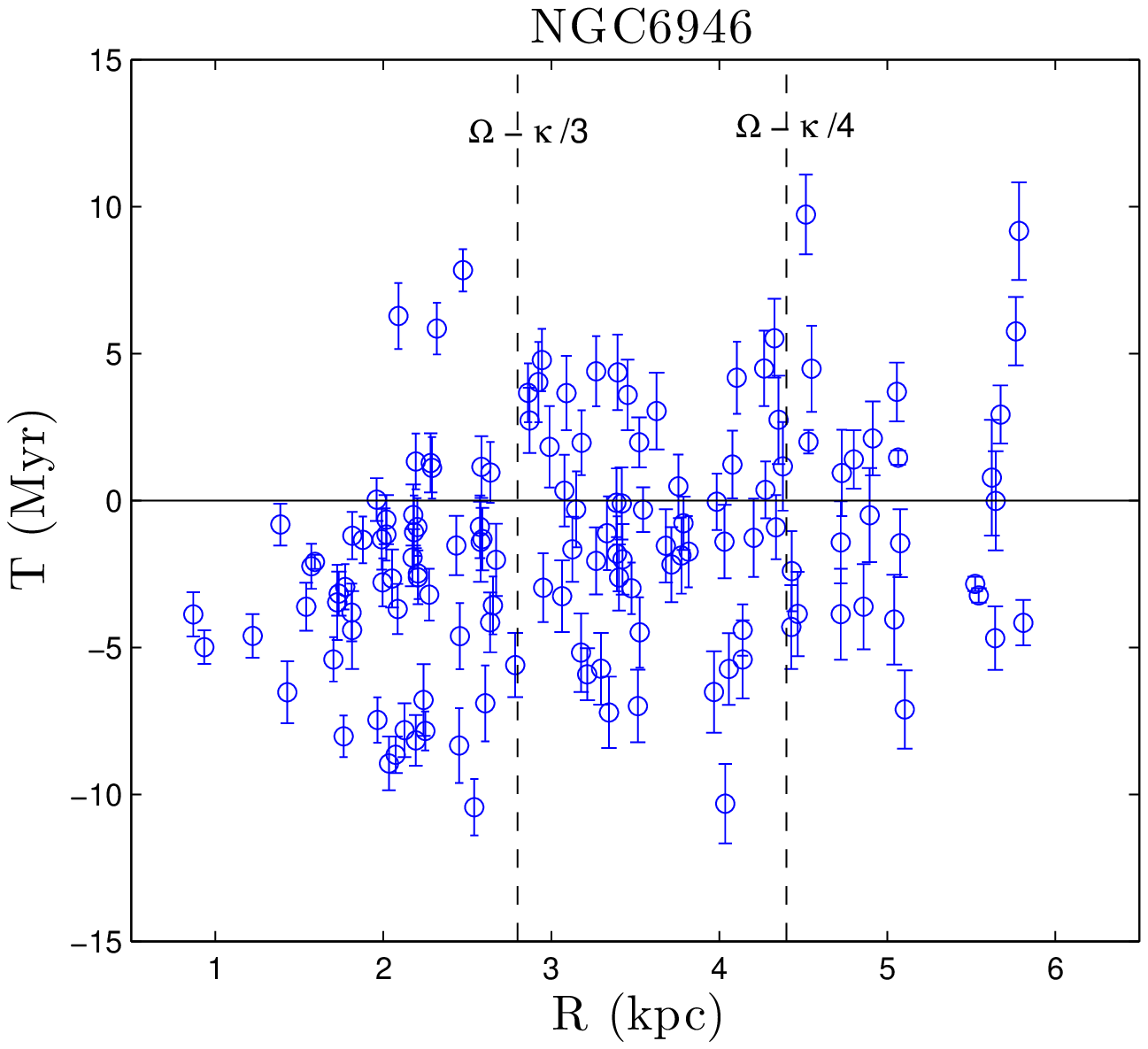}}
  \caption{Parameter $T$ (Eq. \ref{parat}) as a function of galactocentric distance for the sample of selected HII regions (arm regions) of NGC6946.}
  \label{rt69}
\end{figure}

In Figs \ref{rt62}, \ref{rt54} and \ref{rt69} we represent the parameter $T$ as a function of galactocentric distance for the galaxies in the sample. The error bars in these figures were obtained through the error propagation in the equations from the errors in the age of the H~II regions and the error in the determination of the pattern speed, and assuming an uncertainty in the position of the H~II regions equal to the mean seeing of the images. 
The regions closer to $T=0$ were then probably formed by the arm, and those further away were just associated with the arm, but formed in the interarm disc, either before or after the centre of the arm. This effect is clearer for NGC~628 and NGC~5457 (Figs \ref{rt62} and \ref{rt54} respectively), where a small group of regions appears with $T<-20$\,Myr, which are probably formed in the disc before the arm.

\begin{figure}
  \centering
  \resizebox{\hsize}{!}{\includegraphics{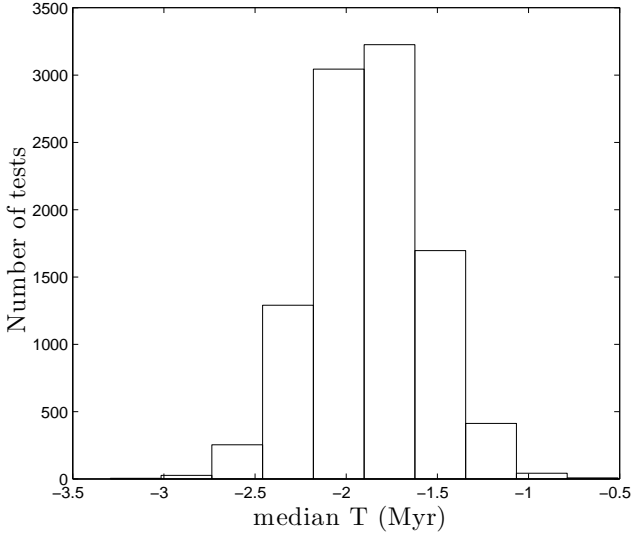}}
  \caption{Histogram with the median values of the parameter $T$ for 1000 tests, for the three galaxies. In each test, the nominal value of the $T$ parameter for each region was varied a random value 3 times its respective error.}
  \label{testtotal}
\end{figure}

It is also apparent that there are more regions with a negative value for $T$ than with a positive value: 79\% of the regions for NGC~638, 57\% of the regions for NGC~5457, and 69\% of the regions for NGC~6946.
Taking into account the uncertainties in the determination of the $T$ parameter, it is possible that the excess of regions with $T<0$ is due to a pure random effect of the errors in the different parameters, and not a significant behavior of the H~II regions. With this hypothesis, the real median value of the $T$ parameter should be approximately equal to zero. Taking into account that the $T$ distributions are not normal, we can not use accurately a statistical test to state the statistical significance of the regions with negative values of $T$. So, in order to check this possible effect, we randomly altered the value of $T$ of each region inside three times its error (we assumed that the uncertainties in $T$ have a Gaussian behavior and three times are equivalent to the 99\% confidence level). We repeated this experiment 10000 times and calculated the median value of $T$ for each time. The results are presented in Fig \ref{testtotal}. For all the tests, the median value of $T$ is always negative, so we may conclude that the presence of a majority of H~II regions with $T<0$ are due the nature of the regions and not an effect of the uncertainties in the data.


\subsection{Star formation rate}

From Figs \ref{rt62} to \ref{rt69} it is clear that there are regions where the star formation started around the moment of the passage of the density wave, some just before and some just after, and regions that were associated with the density wave, since one of the selection criteria for the H~II regions was their spatial relation with the spiral arm, but probably initiated by any other event outside the spiral arms, as will be discussed further on.

In order to explore whether the star formation is really triggered by the arms, it is necessary to obtain the star formation rate of the H~II regions.
To this aim, we will employ the expression derived by Kennicutt (1998b) with the same assumptions we have employed so far: solar abundance and a Salpeter IMF:
\begin{equation}
\left(\frac{SFR}{{\rm M}_{\odot}/{\rm year}}\right) = 7.9\times10^{-42}\left(\frac{L(H\alpha)}{{\rm erg}/{\rm s}}\right),
\label{sfrha}
\end{equation}
assuming case B for recombination, an electron temperature of 10,000 K and no photon leakage.

\begin{figure*}
  \centering
  \begin{tabular}{c c}
    \includegraphics[width=7.5cm]{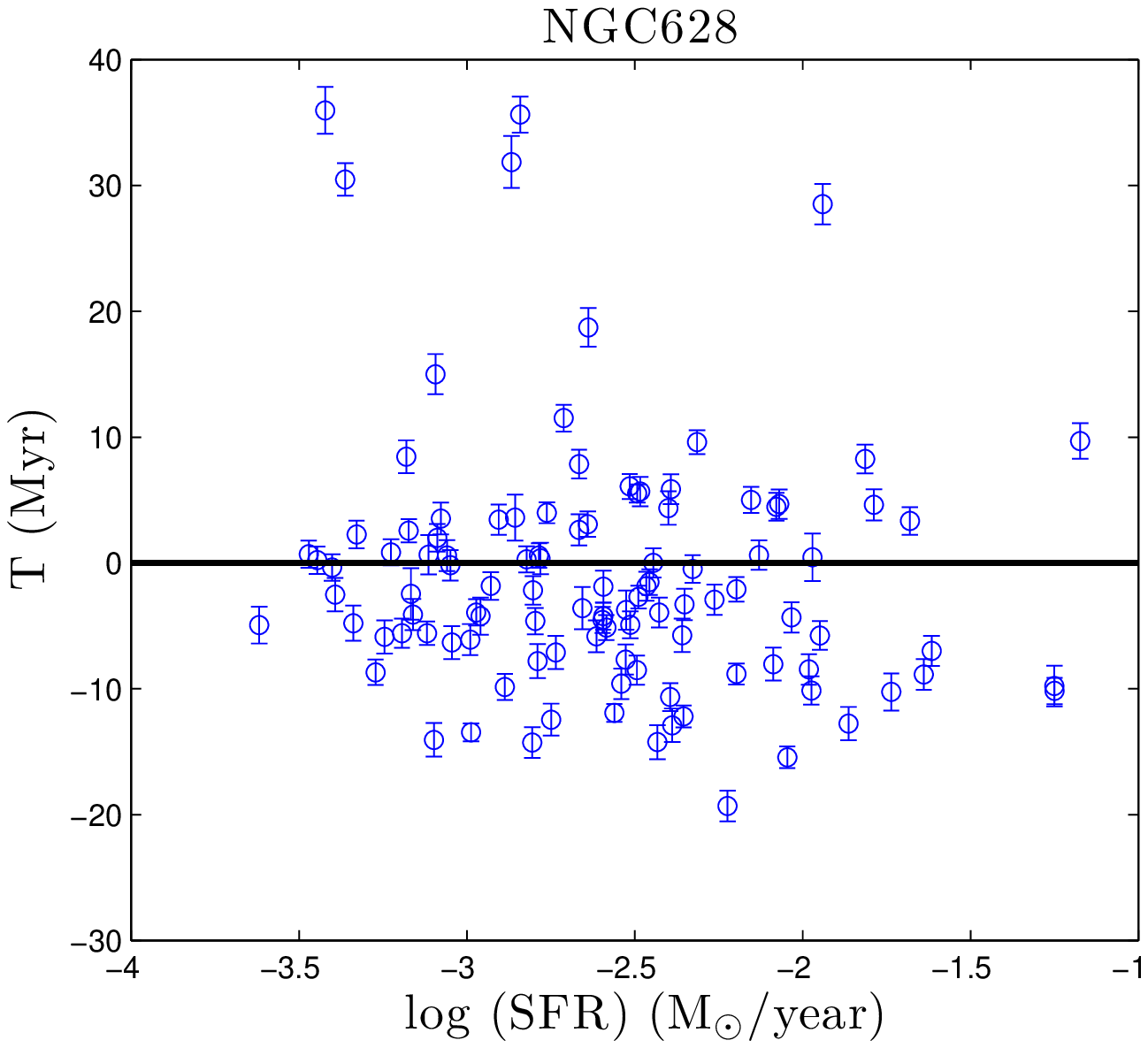} & \includegraphics[width=7.5cm]{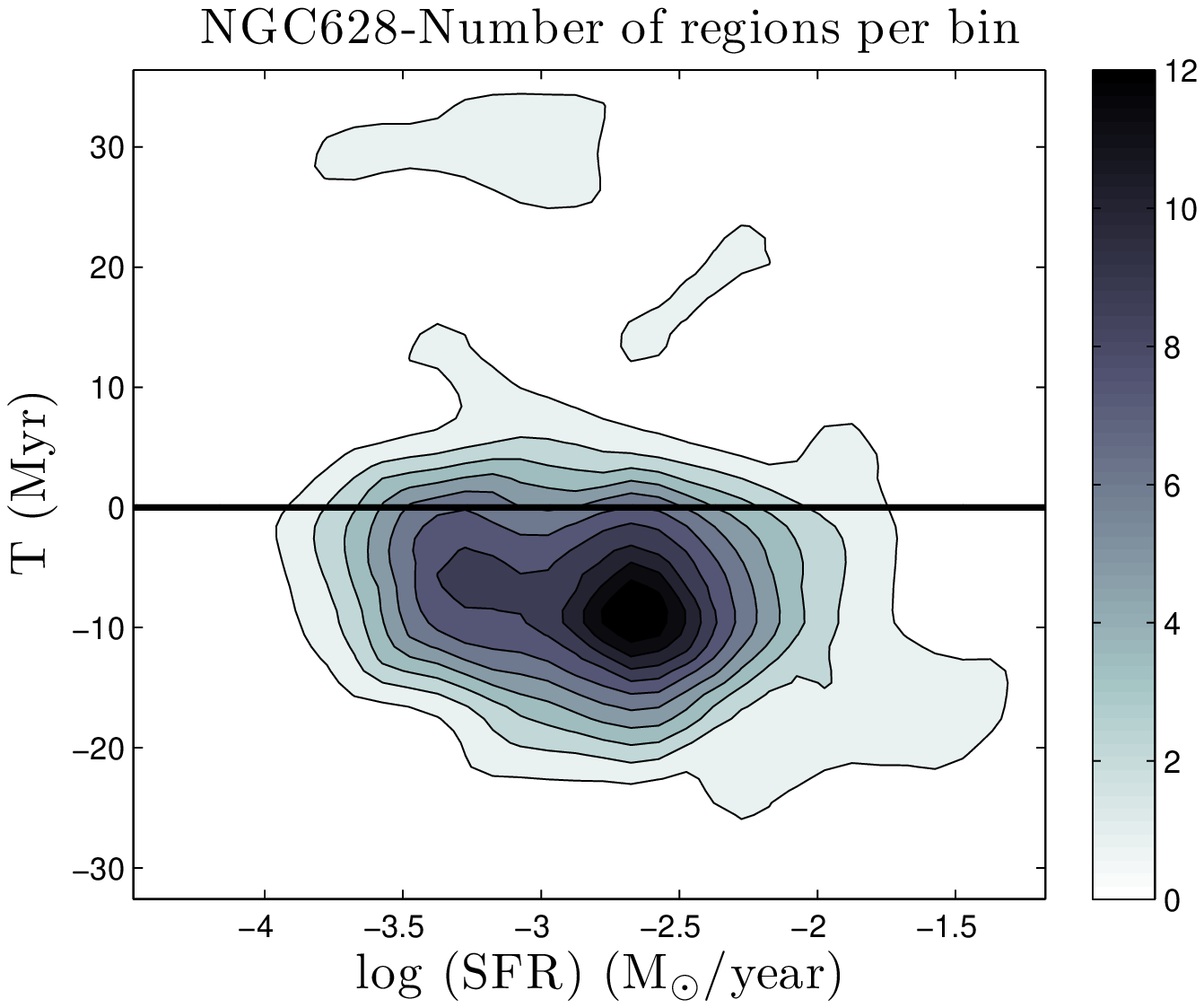} \\
    \includegraphics[width=7.5cm]{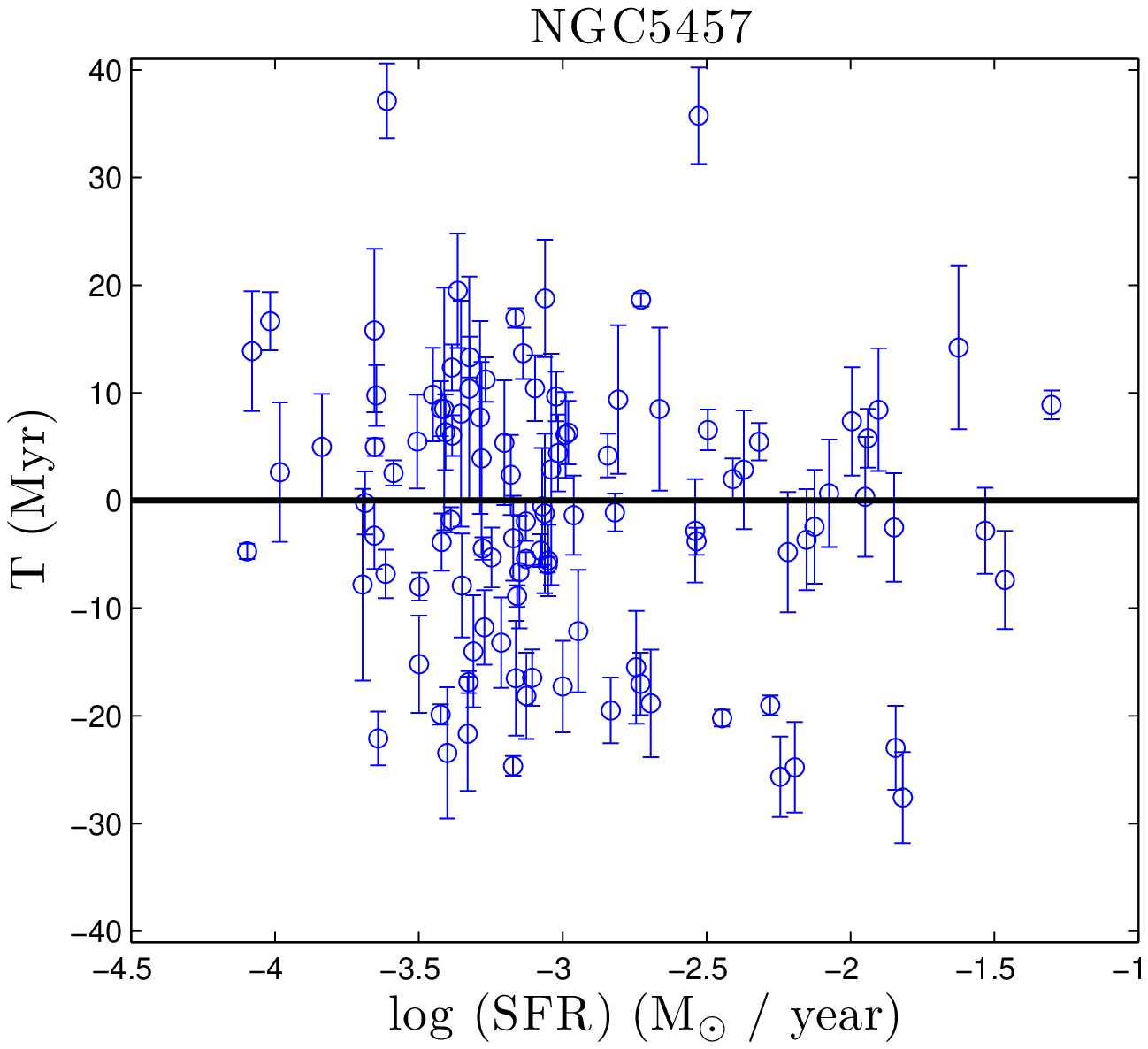} & \includegraphics[width=7.5cm]{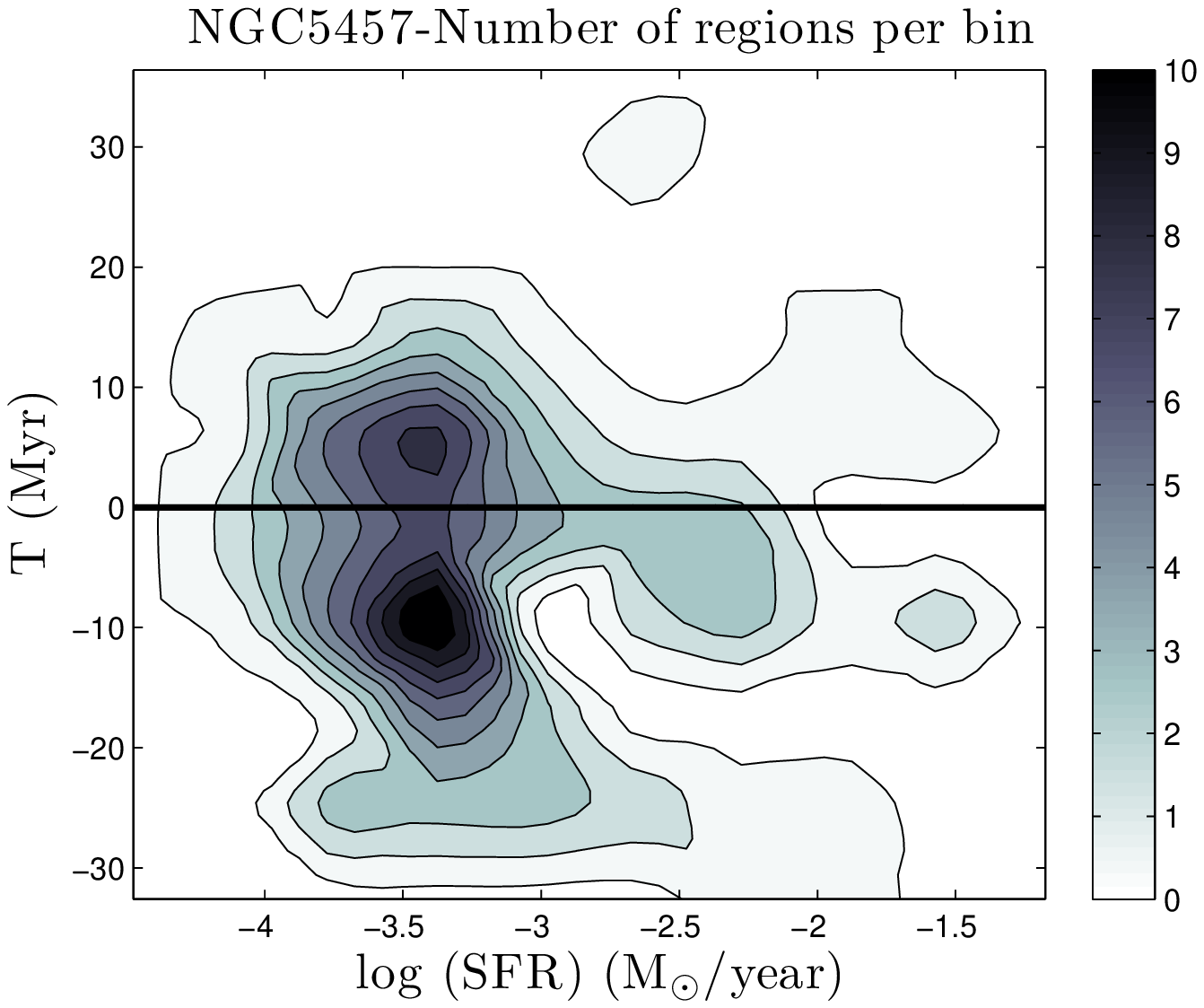} \\
    \includegraphics[width=7.5cm]{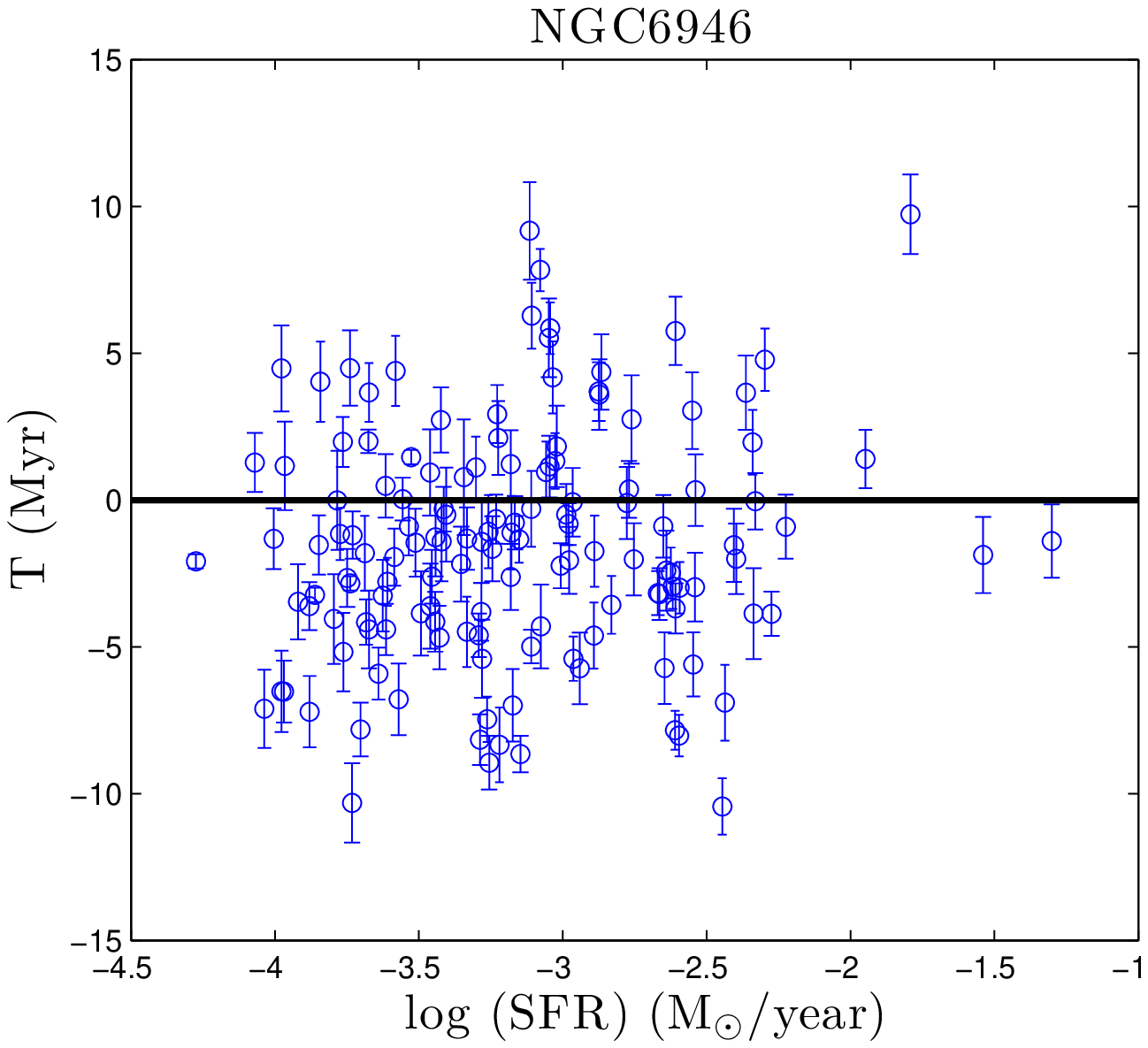} & \includegraphics[width=7.5cm]{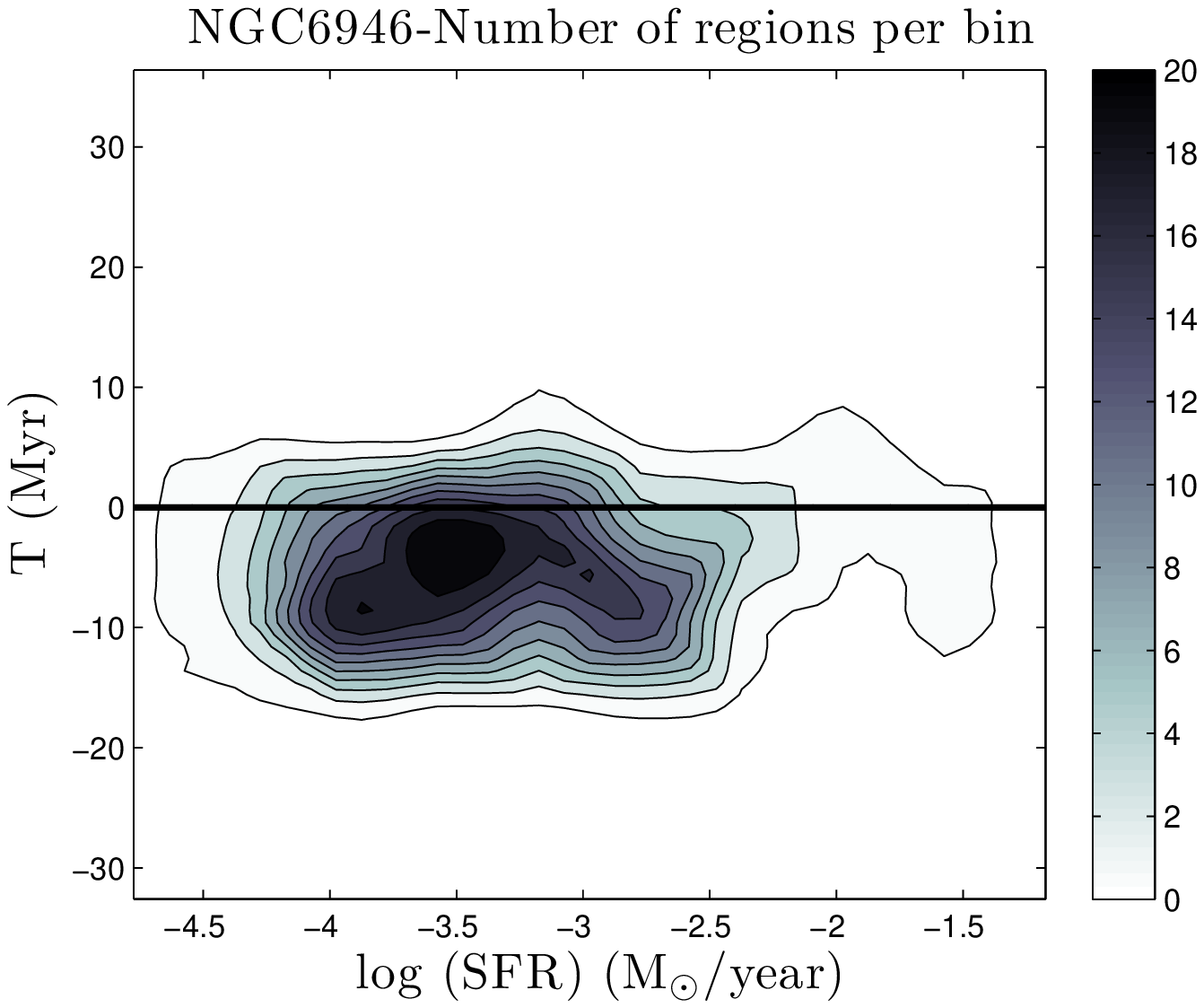} \\
    \includegraphics[width=7.5cm]{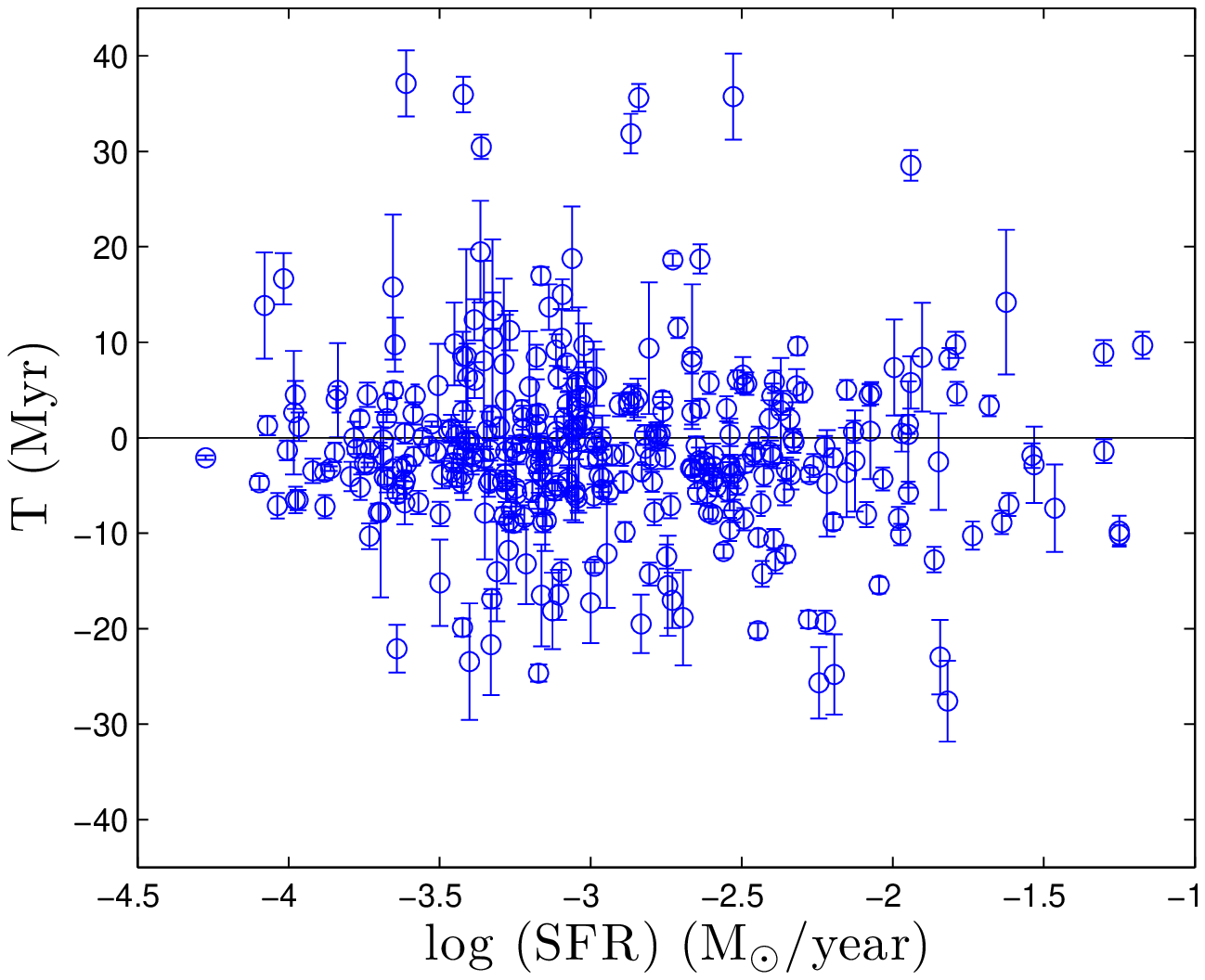} & \includegraphics[width=7.5cm]{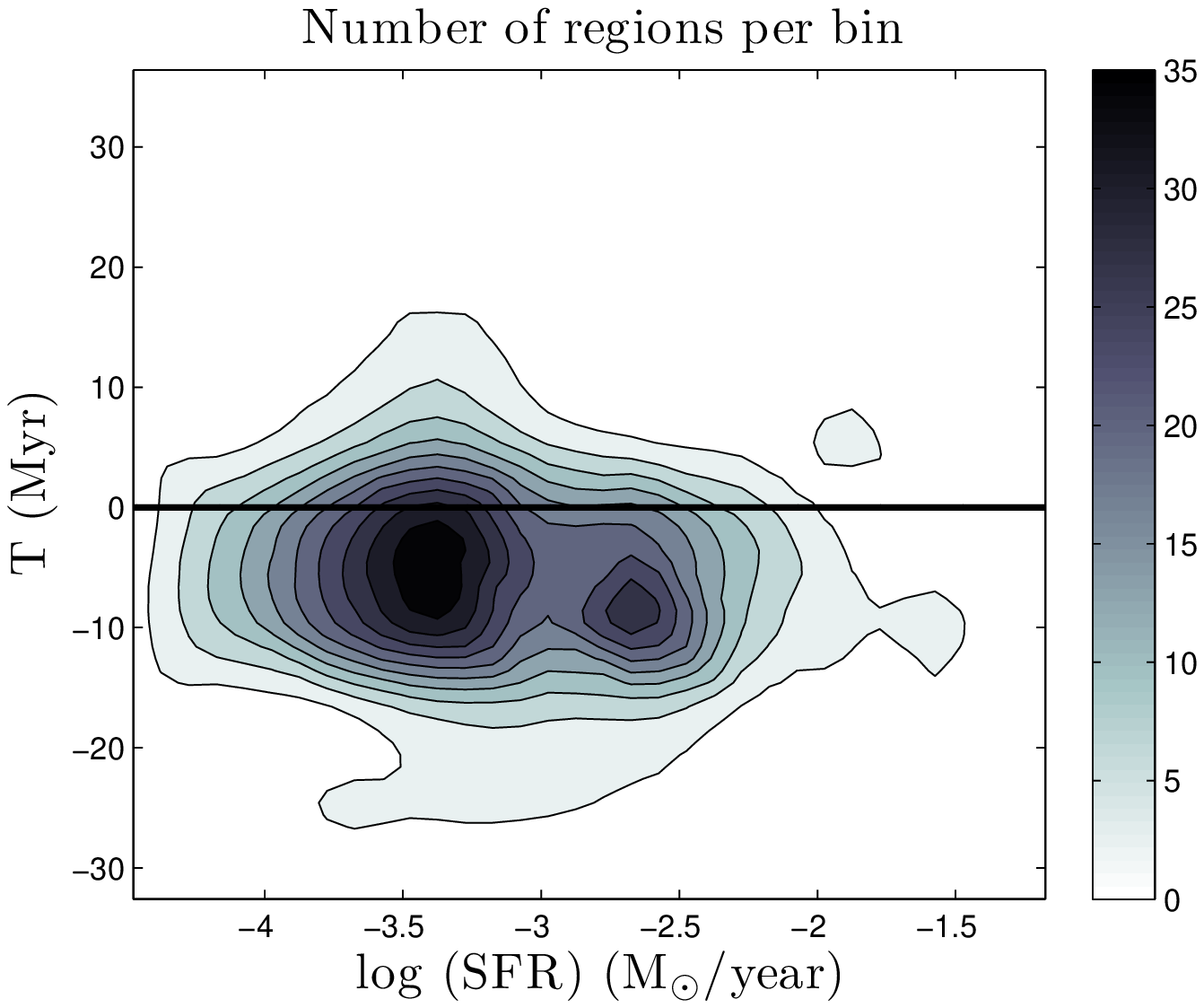} \\
  \end{tabular}
  \caption{Parameter $T$ as a function of the logarithm of the SFR for the H~II regions in our sample (left panels) and the contour plot of the same quantities obtained with bins of 0.4 in $\log (SFR)$ and of 7.7 in $T$ (right panels). The data from the regions of NGC~628, NGC~5457 and NGC~6946 are represented in the uppermost, second and third panels, respectively. The whole sample of H~II regions selected in this study is represented in the lowermost panel. In the right panels, the number of regions per bin is indicated by the colour bars.}
  \label{cont}
\end{figure*}

In Fig. \ref{cont} we have represented the $T$ parameter as a function of $\log (SFR)$ for each galaxy separately and for the total number of H~II regions selected for the three galaxies (left panels). There seems to be no correlation between $T$ and the SFR. However, it is clear that a tendency exists for the H~II regions to be located at $T<0$.
To see this effect more clearly, we have represented in the right panels of Fig. \ref{cont} the contours of the number of regions in bins of 0.4 in $\log (SFR)$ and  7.7\, Myr in $T$ (this divides the $\log (SFR)$ -- $T$ space into a 10$\times$10 reticule and those values for the bins are about three times larger than the mean error for each variable).
The zone with the maximum number of regions is always below the $T=0$ line. For NGC~5457 (second panel from the top), the distribution seems more complex when compared with the rest, with two lobes, one for positive values of $T$ and one (more prominent) for negative values of $T$. For NGC~628 (uppermost panel) and NGC~6946 (third panel from the top), the behavior is more regular, with the maximum in the  $-10<T<0$ zone.
In the contour plot of the total number of regions (lowermost right panel), the bulk of H~II regions are located in a position that it is `behind' (in the direction of the rotation of the galaxy) when compared with the theoretical position they would have if they had been formed in the centre of the arms. This may indicate that the star-forming processes happen some time after the passage of the density wave (indicated by the stellar arm) over the molecular clouds.

\begin{figure}
 \centering
 \resizebox{\hsize}{!}{\includegraphics{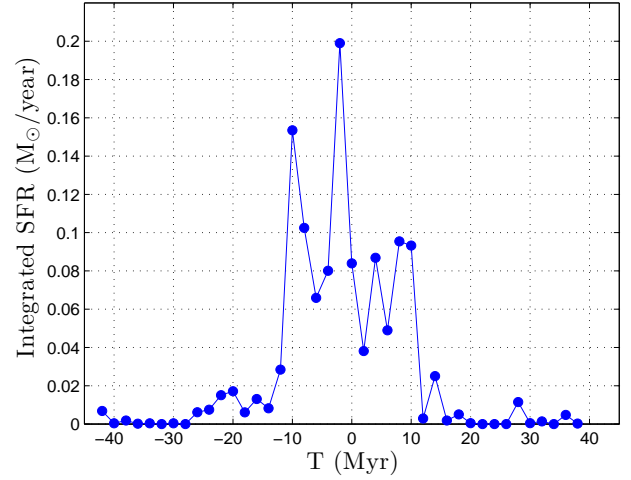}}
 \caption{Integrated SFR in bins of 2\,Myr versus the $T$ parameter for all the H~II regions in the sample.}
 \label{pico}
\end{figure}

In Fig. \ref{pico} the integrated SFR versus parameter $T$ is represented. The SFR has been integrated in bins of 2\,Myr  width.
The largest SFR is concentrated just after the passage of the central part of the arm and there is a large fall in the value of the SFR (about 80\%) just for the regions at $T=0$.
The peaks in the distribution of the data are below the line at $T=0$, and this could indicate that not only were those regions triggered by the arm, but that they were also created after the passage of the density wave. This, together with the distribution of the regions presented in Fig.\ \ref{cont}, seems to agree with the scenario proposed by Roberts (1969), in which star formation occurs after the passage of the density wave. In order to confirm this behavior, more observations of different galaxies with different arm classes are required.

\section{Evidence for triggering of star formation}

Even taking into account that there is a percentage of H~II regions formed just after the passage of the density wave, we have yet to find clear evidence for the triggering of star formation. We consider that there is triggering when the SFR in the arm is much larger than the SFR value in the interarm times the density ratio between both zones (Cepa \& Beckman, 1990).
In an analogous method to that proposed by Cepa \& Beckman (1990), we define a relative arm efficiency of massive star formation ($\epsilon$). This relative arm efficiency can be described as the ratio of the SFR per area unit between the arm and interarm regions at the same radial distances divided by the amplitude of the arm measured in I-band images. This amplitude is defined as the ratio of the I flux in arbitrary units per area unit between the arm and interarm at similar angular and radial distances (Elmegreen \& Elmegreen 1984), and can be considered a lower boundary of the density wave amplitude, since the amplitude of the stellar density wave is lower than that of the gas (Elmegreen \& Elmegreen 1985; Elmegreen et al.\ 1992). We have done this in radial bins of 1\,kpc for the galaxies NGC~628 and NGC~5457, and 0.5\,kpc for NGC~6946 (because this galaxy is closer to us than the other two, see Table \ref{param}). Therefore, the efficiency at radius $R$ is

\begin{equation}
\epsilon=\displaystyle\displaystyle\frac{\displaystyle\sum_{R}SFR_{arm}}{\displaystyle\sum_{R}SFR_{interarm}}\times\displaystyle\frac{{\displaystyle\sum_{R}I_{interarm}}}{{\displaystyle\sum_{R}I_{arm}}},
\end{equation}
where $SFR_{arm}$ and $SFR_{interarm}$ are the SFR of an H~II region at a radius $R$ per area unit in the arm or in the interarm respectively, and $I_{arm}$ and $I_{interarm}$ are the I fluxes per area unit in arbitrary units at the same radius of the H~II region. The summation is done over all the regions inside the bin whose mean radius is $R$.  
This dimensionless factor takes into account the difference in density between arm and interarm zones, so it is a good indicator of the triggering of the star formation. Indeed, if $\epsilon$ is of the order of  unity, there is no triggering, and if $\epsilon$ is much larger than one, a star formation triggering is happening at the selected radius (Cepa \& Beckman 1990).

To obtain the values of $\epsilon$ we need to calculate the SFR for inter-arm regions. However, the selection criterion employed in this study is based on selecting only those regions closer to the arms, so the possibility of a lack of interarm regions exists. For this reason, in this case we will consider all the regions with $\left|\Delta\theta\right|<100^\circ$ for NGC~628 and NGC~5457, and $\left|\Delta\theta\right|<75^\circ$ for NGC~6946 (with $\Delta\theta$ defined by eq. \ref{deltateta}) as inter-arm regions. With these figures we will include all the closer interarm regions and we will minimize the inclusion of H~II regions from other arms. Due to the large number of arms and satellite arms of NGC~6946 compared with the other two galaxies in the sample, the maximum value selected for $\left|\Delta\theta\right|$ is smaller than the one employed for NGC~628 and NGC~5457 (see Figs \ref{arm628} to \ref{arm6946}).

\begin{figure}
 \centering
 \resizebox{\hsize}{!}{\includegraphics{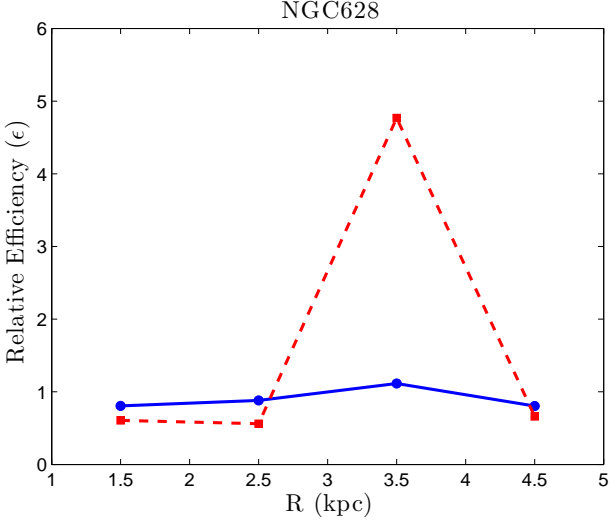}}
 \caption{Relative arm/interarm star formation efficiency for NGC~628 versus the deprojected galactocentric radius in bins of 1\,kpc width. The filled dots and continuous line represent the data from NSN arm, filled squared and dashed line represent the data from SNS arm.}
 \label{epsi628}
\end{figure}

In Fig. \ref{epsi628} we have represented the relative efficiency as a function of the deprojected galactocentric radius for the two arms of NGC~628. For both arms, the efficiency is rather low (of the order of unity)  along the entire length of the radius. However, for the SNS arm, in the bin with its centre at 3.5\,kpc there is a relative large increase in the efficiency, indicating a possible triggering of star formation. This triggering is only of massive star formation, since H$\alpha$ is sensitive to ionizing stars, and the IMF might be biased in the arms in the sense of a larger fraction of massive stars in the arms than in the inter-arm regions (Cedr\'es et al.\ 2005). Cepa \& Beckman (1990) also found  triggering, but closer to the centre of the galaxy (at about 2\,kpc). Nevertheless, it has to be taken into account that the method of calculating the relative efficiency by Cepa \& Beckman (1990), the distance employed and the inclination angle are all different to those used in this study. 

\begin{figure}
 \centering
 \resizebox{\hsize}{!}{\includegraphics{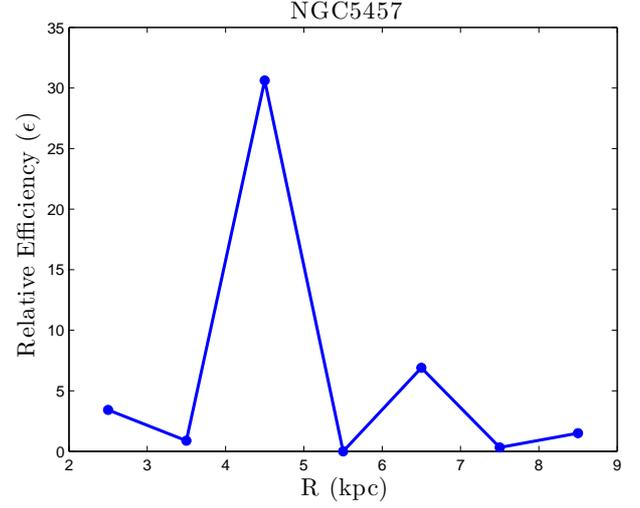}}
 \caption{Relative arm/interarm star formation efficiency for NGC~5457 versus the deprojected galactocentric radius in bins of 1\,kpc width.}
 \label{epsi5457}
\end{figure}

In Fig.\ \ref{epsi5457} we have represented the relative efficiency as a function of the deprojected galactocentric radius for NGC~5457. This galaxy presents two noteworthy peaks at bins centered at$ R=4.5$\,kpc and $R=6.5$\,kpc. The first one indicates an important episode of triggering of star formation activity in the arm compared with the interarm. Another remarkable feature for this galaxy is the $\epsilon=0$ at $R=5.5$\,kpc.

\begin{figure}
 \centering
 \resizebox{\hsize}{!}{\includegraphics{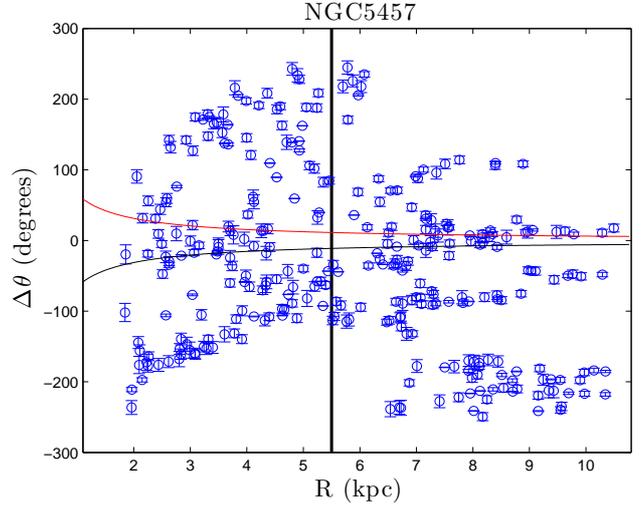}}
 \caption{Angular distance $\Delta\theta$ as obtained in eq.\  \ref{deltateta} versus the deprojected galactocentric radius for all the regions in NGC~5457. The vertical line indicates the suggested position for the co-rotation radius. The other two curves indicate the angular limits of the arm.}
 \label{teta54}
\end{figure}

To study this complete lack of efficiency, in Fig.\ \ref{teta54} we have represented the value of $\Delta\theta$ for all the regions detected in the galaxy versus galactocentric radius. The two curves represent the limits of the arm studied. There is an apparent lack of H~II regions formed in the arm from 4.75\,kpc to 6\,kpc. The vertical line marks the radius where the measured relative efficiency is 0 in Figure \ref{epsi5457}. Moreover, in Fig.\ \ref{rt62} is observed the lack of H~II regions between 5 to 6\,kpc. This lack of regions formed by the passage of the density wave may indicate the presence here of the co-rotation radius, an effect already suggested in Cepa \& Beckman (1990). Indeed, Egusa et al.\  (2009) give a range for the position of the co-rotation from 2\,kpc to 6.7\,kpc in radius, which is compatible with this result. This radius is indicated in Fig.\ \ref{armm101} by the dot-dashed line. Moreover, as noted in del R\'{\i}o \& Cepa (1998), the presence of a bifurcation of the arm studied about this radius increases the plausibility of the position of co-rotation near this location. However, this galaxy may have several pattern speeds, (see, for example Meidt et al. 2009), so it is possible that some overlapping exists between the co-rotation of the inner disc pattern and some resonance for one of the outer patterns (according to Meidt et al. 2009, this could be $\Omega-\kappa/2$ for a pattern of $\Omega=18$\,km\,s$^{-1}$\,kpc$^{-1}$), that may explain the lack of regions in this arm at the mentioned galactocentric radius.

\begin{figure}
 \centering
 \resizebox{\hsize}{!}{\includegraphics{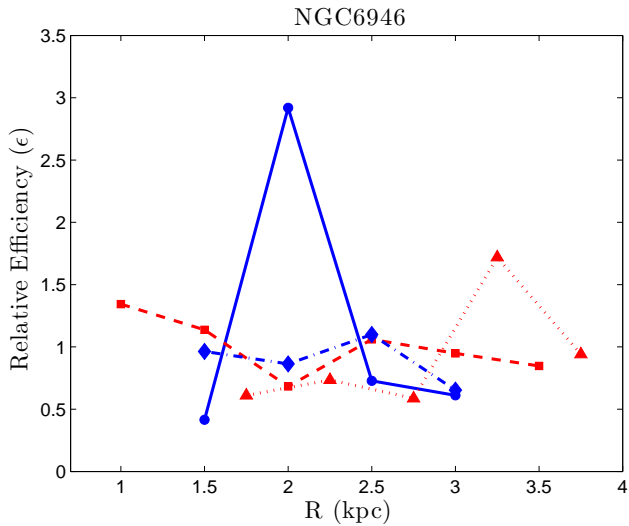}}
 \caption{Relative arm/interarm star formation efficiency for NGC~6946 versus the deprojected galactocentric radius in bins of 0.5\,kpc width. Symbols as in Figure \ref{arm6946}}
 \label{epsi6946}
\end{figure}

In Fig. \ref{epsi6946} the relative efficiency as a function of the deprojected radius for NGC~6946 is shown. Star formation triggering does not seem to exist for all the arms of this galaxy, but just a small increase for one of the arms at R=2\,kpc. This is not surprising, because: a) there is a lack of interarm H~II regions and b) even if this galaxy is classified as class 9, according to Elmegreen \& Elmegreen (1987), it presents a more chaotic structure of the arms when compared with NGC~628 and NGC~5457, and it can be considered as more flocculent than the other galaxies of the sample (Foyle et al.\ 2010). Fathi et al.\ (2007), situates the co-rotation radius for this galaxy at 8.3\,kpc.  

\section{Conclusions}
We have selected H~II regions, and associated them by proximity with several spiral arms of the grand design galaxies NGC~628, NGC~5457 and NGC~6946. For those regions, we were able to obtain their ages, from the H$\alpha$ equivalent width, employing Leitherer et al.\ (1999) models, assuming a Salpeter IMF and a solar abundance.

Using the method described in Egusa et al.\ (2009), we calculated the positions in the disc that the arm H~II regions should have if they had been formed at the maximum of the density wave. These theoretical positions were compared with the real positions through a new parameter, $T$, defined as the time required for an H~II region to move the offset between its real position and the position of a hypothetical region formed in the center of the arm. This parameter is independent of the radius of the galaxy.

We have calculated the SFRs for the selected H~II regions and we have represented them as a function of the $T$ parameter. We found a concentration of H~II regions in $-10<T<0$. Moreover, the bulk of the integrated SFR is also in the $-10<T<0$ range, in agreement with Roberts (1969) suggesting that star formation is taking place after the passage of the density wave. 

The relative efficiency of the SFR between arm and interarm zones as a function of the galactocentric radius was calculated. Evidence of considerable triggering of star formation was found for NGC~5457 and a possible position for the co-rotation radius is suggested to be about R=5.5\,kpc, where there is a lack of star formation regions. However, the spiral structure of NGC~5457 is very complex and this lack of presence of H~II regions and low star formation efficiency in the zone may be due the overlapping of multiple pattern speeds that may exist for this galaxy. A more modest evidence of triggering was found for NGC~628 for one of the arms. The wide range of uncertainties for the position of the co-rotation given by Egusa et al.\ (2009) makes it difficult to give a firm conclusion concerning the low efficiency at $R=2.5$\,kpc, just before the small triggering at 3.5\,kpc. The results seem to suggest a absence of triggering of star formation for NGC~6946. However, it should be noted that this galaxy is more flocculent than the other two. On the other hand, NGC~6946 presents a clear lack of H~II regions not generated in the arms that may introduce a bias in the calculation of the efficiency.

\begin{acknowledgements}
This work was supported by the Spanish Ministry of Economy and Competitiveness (MINECO) under grants AYA2011-29517-C03-01 and AYA2010-08896-E.
It is also a pleasure to thank the anonymous referee for very constructive comments and suggestions that helped to greatly improve the paper.
\end{acknowledgements}

\end{document}